# Competition between Induced-Charge Electro-Osmosis and Electro-Thermal Effects around a Weakly-Polarizable Microchannel Corner

Matan Zehavi, Alicia Boymelgreen and Gilad Yossifon*

*Faculty of Mechanical Engineering, Micro- and Nanofluidics Laboratory, Technion–Israel Institute of Technology, Technion City 32000, Israel*

The microchannel corner is a common inherent component of most planar microfluidic systems and thus its influence on the channel flow is of significant interest. Application of an alternating current electric field enables quantification of the non-linear induced-charge electro-osmosis (ICEO) ejection flow effect by isolating it from linear electro-osmotic background flow which is present under dc forcing. The hydrodynamic flow in the vicinity of a sharp channel corner is analyzed using experimental micro-particle-image-velocimetry (μPIV) and numerical simulations for different buffer concentrations, frequencies and applied voltages. Divergence from the purely ICEO flow with increasing buffer conductivity is shown to be a result of increasing electro-thermal effects due to Joule heating.

*Corresponding author: yossifon@tx.technion.ac.il

## I. INTRODUCTION

Previous experimental studies of induced-charge-electro-osmosis (ICEO) around weakly polarizable (i.e. dielectric) structures embedded within an electrolyte have focused mainly on the direct current (DC) case [1]–[7]. At the same time, a number of theoretical studies of time dependent ICEO have been performed for perfectly conducting cylinders in both step-wise and alternating current (AC) fields [8], lossless dielectric structures using macroscale time-dependent boundary conditions [9], lossy dielectric cylinders under AC fields [10] and dipolophoresis of Janus particles [11]. In contrast to "alternating-current electroosmosis" (ACEO) which describes hydrodynamic flow generated by polarization of the active electrodes themselves[12], ICEO, whether under DC or AC fields, refers to a floating structure which is polarized by a field applied at external electrodes[8].

In the current work we focus on ICEO processes occurring at a sharp corner of a L-shaped microchannel junction. This work extends previous studies [5]–[7] focused on colloid and fluid flow dynamics in the same microchannel design, but utilizes AC rather than DC applied fields. In contrast to the DC case where the effects of natural and induced zeta potentials are superimposed, in the AC case, only the latter non-linear effect that has a non-vanishing time-averaged effect. In addition it will be shown that Joule heating resulting in electro-thermal effects [13] becomes significant at sufficiently large solution conductivity/applied electric fields. In the following, we describe the experimental methods in sec. II, the numerical model in sec. III, the results and discussion in sec. IV and concluding comments in sec. V.

## II. EXPERIMENTAL METHODS

### A. Fluidic device and experimental setup

The fluidic device is made of two reservoirs connected by a L-junction micro-channel, (Figure 1) 140µm in depth (normal to the plane of view). The micro-channels were fabricated from PDMS (Polydimethylsiloxane, Dow-Corning Sylgard 184) using the rapid prototype technique [14]. A high resolution chrome mask with a 3µm resolution was used for the SU8 mold. The nominal L-junction corner has a radius of curvature is 6±2 µm. The channel was bonded to a microscope slide coated with a 30µm layer of PDMS using plasma bonding[15]. An electric function generator (Agilent 33250A) was connected through an amplifier (TREK 2220) to platinum wire electrodes (0.5mm platinum wire, Sigma-Aldrich) which were inserted in the reservoirs. The electric signal was monitored using an oscilloscope (Tektronix TPS-2024).

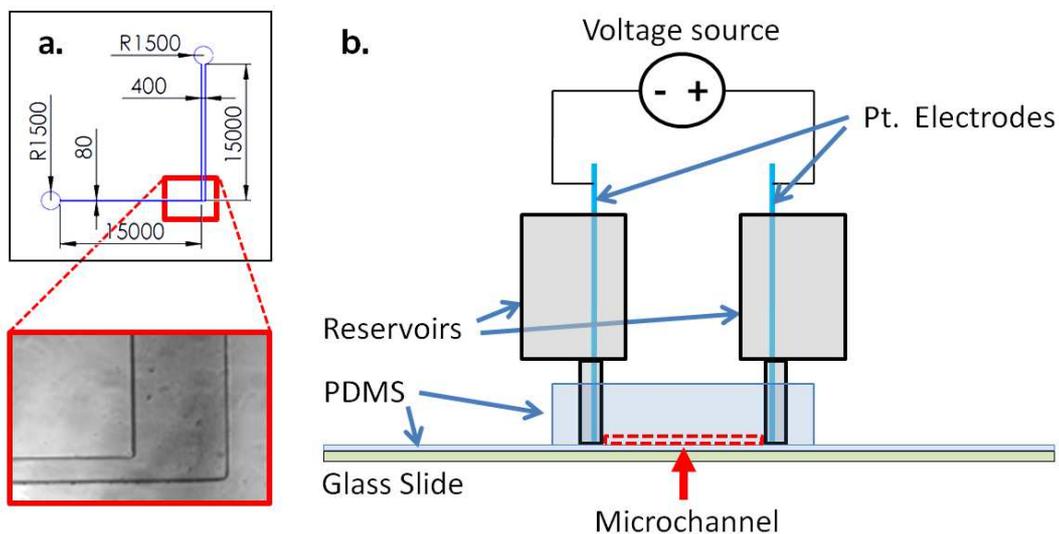

**Figure 1**: **(color online)** (a) Schematic design of the microfluidic chip and scaled up microscopic image of the L-junction; (b) experimental setup (not to scale). All dimensions are in µm.

### B. μ-PIV measurements

KCl solutions of 3 different concentrations (solution conductivities of 2.21±0.01, 50.6±0.4, 373.0 μS/cm) were seeded with tracer particles (Thermo Scientific 0.48μm red Fluoro-Max Fluorescent Particles). Particle imaging was performed using a commercial μPIV system (TSI) installed on a Nikon TI microscope. The PIV images were obtained with a Litron Nano S65-PIV Laser (YAG 532nm wavelength, 400mJ, 4ns at 15Hz) through a Nikon Plan-Fluor 20x/0.50 and a dichroic filter (560nm dichroic/565nm emission). The time interval between images was 10ms, with a repetition rate of 5 image pairs per second, and 50 image pairs per measurement set. The images were then passed through a linear Gaussian filter (Filter size 5x5 pixels, standard deviation parameter 0.5) and background subtraction. The processing interrogation region is 32x32 pixels on a standard Nyquist grid, with the results presented as a sum of correlations of the 50 image pairs (see [16] for more details on μPIV technique).

### C. Temperature visualization

Ross et al. [17] developed a Rhodamine B based thermometry technique for direct measurements of the in-channel temperature profiles in microfluidic systems. Rhodamine B is a fluorescent dye whose quantum yield is strongly dependent on temperatures in the range of 0°C to 100°C, making it ideal for liquid based systems. Here, the neutral (zwitterionic) fluorescent dye Rhodamine B (Sigma 83690) at $10^{-4}$M in concentration was added to KCl solutions of different conductivities 2.1, 50.3, 351μS/cm, corresponding to concentrations of $1.4 \times 10^{-5}$, $3.3 \times 10^{-4}$, $2.3 \times 10^{-3}$M, respectively. The chip was placed on a Nikon TI microscope and illuminated using a fluorescent lamp (Nikon-Intensilight C-HGFL, wavelength 380-600nm, 130Wx50Hz). Images were captured using Andor-NEO camera, through a Nikon Plan-Fluor 10x/0.30 objective lens and an Omega Optical (560nm dichroic/565nm emission) dichroic filter. Ambient temperature of $24°C$ was measured on the glass slide surface using a K-type thermocouple.

## III. NUMERICAL MODEL
### A. Electrostatics

The AC electrostatic problem for the channel geometry shown in Figure 1(a) is solved numerically using finite-element based commercial software (COMSOL). Herein, we follow the approach taken in Yossifon 2009 [9] wherein macroscale boundary conditions for the time-dependent problem (eqs. 2.15, 2.16 in [9]) were derived in a way that eliminated the need to directly resolve transport within the electric double layers (EDL). A symmetric electrolyte ($z^+ = -z^- = z$) with equal ionic diffusivities ($D^+ = D^- = D$) is considered along with the standard assumptions of a thin EDL ($\delta = \lambda/a \ll 1$) and weak electric fields ($\psi = zF\phi_0/RT \ll 1$). Herein, $\lambda$ denotes the Debye length, $a$ the geometric length scale of the problem chosen to be the corner radius of curvature, $F = 9.648 \cdot 10^4 \, Cmol^{-1}$ is the Faraday number, $R = 8.314 \, JK^{-1}mol^{-1}$ the universal gas constant, $T = 300°K$ the absolute temperature and $\phi_0$ the characteristic induced electric potential. The latter was set [9] to $\phi_0 = E_0 \varepsilon_w \lambda a / (\varepsilon_f a + \varepsilon_w \lambda)$, wherein $\varepsilon_f$ and $\varepsilon_w$ denote the dielectric constants of the electrolyte solution and solid wall, respectively and $E_0$ the characteristic magnitude of the external field within the narrow microchannel. For the current geometry, $E_0 = \left(\frac{\Delta V}{L}\right)\left(\frac{h_2}{h_1+h_2}\right) \approx 1.1 \cdot 10^5 \, V/m$, where $h_1 = 80\mu m$, $h_2 = 400\mu m$, $L = 15mm$ so that for $\Delta V = 2000V$, $\phi_0 \approx 5mV$ in rough agreement with the weak field assumption.

Within this framework, the hydrodynamic and electrostatic problems are decoupled. Furthermore, the electrostatic solution can be separated into three sub-domains: the 'outer' electroneutral solid wall and bulk solutions with their harmonic electric potentials $\phi_w$ and $\phi_f$, respectively, and the 'inner' domain of the diffuse EDL whose electric potential $\phi$ satisfies Poisson's equation. The macroscale boundary conditions connecting $\phi_w$ and $\phi_f$ on the solid-electrolyte interface, constituting the main result of [9], were obtained by focusing on the inner domain. The linearity of the present problem allows for the superposition of the linear equilibrium (exclusively associated with a native surface charge/zeta potential) and nonlinear induced charge solutions. However, when using AC electric forcing, the former linear solution integrates to zero over a given period so that in the sequel we need focus only on the induced-charge distribution.

For the special case of harmonic AC forcing, i.e. $E = e^{j\omega t}$ within the narrow channel away from the junction and normalized by $E_0$, the potentials within the solid wall and bulk solution can be expressed as $\phi_f = \underline{\phi}_f e^{j\omega t}$ and $\phi_w = \underline{\phi}_w e^{j\omega t}$, respectively. Herein, $\underline{\phi}_f$ and $\underline{\phi}_w$ are the non-dimensional complex amplitudes (phasors) normalized by $\phi_0$, $j$ is the imaginary number, $\omega$ is the non-dimensional frequency normalized by $1/\tau$ and $t$ is the time normalized by the EDL diffusion time $\tau = \lambda^2/D$. Both potentials obey the Laplace equation

$$\nabla^2 \underline{\phi}_f = 0, \quad \nabla^2 \underline{\phi}_w = 0, \quad (1)$$

along with boundary conditions

$$\underline{\phi}_f - \underline{\phi}_w = \frac{\alpha}{\gamma^3} \frac{\partial \underline{\phi}_w}{\partial n}, \quad (2a)$$

$$\frac{\alpha}{\delta} \frac{\partial \underline{\phi}_w}{\partial n} = \left(\frac{\gamma^2}{\gamma^2 - 1}\right) \frac{\partial \underline{\phi}_f}{\partial n}, \quad (2b)$$

which were directly derived from Eqs. 2.15', 2.16' [9] for the special case of harmonic forcing in the limit of infinite time (i.e. when the transient phase of the solution vanishes to obtain a periodic solution). Herein, $\gamma^2 = 1 + j\omega = 1 + \frac{\lambda^2}{D} j\tilde{\omega} = 1 + \frac{\lambda^2}{D} j 2\pi \tilde{f}$, $\tilde{f}$ is the frequency and $n$ is the spatial coordinate normal (in the outward sense) to the surface of the solid, normalized by $a$. The tilde stands for dimensional quantities. The parameter $\alpha = (\varepsilon_w/\varepsilon_f)\delta$, in the terminology of an equivalent RC-circuit model, represents the ratio of the capacitance of the dielectric solid $\sim \varepsilon_w/a$ to that of the EDL $\sim \varepsilon_f/\lambda$. Eqs.(2a,b) also satisfy the extended boundary conditions of Zhao and Yang[10] in the limit of an ideally dielectric wall (i.e. lossless dielectric), which for the case of PDMS used in the current study (material conductivity O($10^{-10}$μS/cm[18]), constitutes a reasonable assumption.

We note that substitution of (2a) into (2b) yields the charging equation for the EDL

$$\underline{\zeta} = \underline{\phi}_w - \underline{\phi}_f = -\frac{\delta}{\gamma(\gamma^2 - 1)} \frac{\partial \underline{\phi}_f}{\partial n}, \quad (3)$$

where $\underline{\zeta}$ is the induced part of the zeta potential phasor in terms of the electric potential in the fluid domain. For the DC case $(\gamma = 1)$ eqs. (2a,b) reduce to

$$\frac{\partial \underline{\phi}_f}{\partial n} = 0, \quad (4a)$$

$$\underline{\phi}_w + \alpha \frac{\partial \underline{\phi}_w}{\partial n} = \underline{\phi}_f, \quad (4b)$$

which is in agreement with previous results [5], when the contribution of the equilibrium zeta potential is neglected.

For the sake of comparison with other related studies, the above results can be examined within certain limits. For example, in the case of a conductive cylinder (i.e. $\underline{\phi}_w = 0$) subject to an externally applied field $e^{j\omega t}\hat{\mathbf{z}}$ (normalized by $E_0$, wherein $\hat{\mathbf{z}}$ is the normal unit vector of the axisymmetry axis of coordinate $z = r\cos(\theta)$), the solution for the potential phasor is of the form [8]

$$\underline{\phi}_f = -(E_0 a/\phi_0) r \cos(\theta)\left[1 + g_{(\omega)}/r^2\right], \quad (5)$$

with the coordinate $r$ normalized by the cylinder radius $a$ and $g = (1 - j\omega\delta^{-1})/(1 + j\omega\delta^{-1})$. From eqs. (3) and (5) one obtains the induced zeta potential phasor

$$\underline{\zeta} = 2\left(\frac{E_0 a}{\phi_0}\right)\frac{\cos(\theta)}{\gamma(1 + j\omega\delta^{-1})}, \quad (6)$$

which for small enough frequencies (i.e. $\omega \ll 1$), i.e. $\gamma^2 = 1 + j\omega \cong 1$, stands in agreement with eq. 4.9 in Squires & Bazant [8].

**B.    Thermal analysis**

For sufficiently high solution conductivity, $\sigma$ [$Sm^{-1}$], and/or large electric fields, Joule heating effects [19], [20] may become significant due to generation of a large power density within the fluid domain

$$\dot{q} = \left\langle \sigma |\tilde{\mathbf{E}}|^2 \right\rangle \quad [Wm^{-3}], \quad (7)$$

with $\sigma$ being the solution conductivity, $\tilde{\mathbf{E}} = -\tilde{\nabla}\tilde{\phi}_f$ is the electric field within the fluid domain and $\langle \ \rangle$ the time average. The temperature field is obtained by solving the energy equation with the internal Ohmic heat source $\dot{q}$

$$\rho c_p \frac{D\tilde{T}}{D\tilde{t}} = \rho c_p \left(\frac{\partial \tilde{T}}{\partial \tilde{t}} + \tilde{\mathbf{u}} \cdot \tilde{\nabla}\tilde{T}\right) = k_f \tilde{\nabla}^2 \tilde{T} + \dot{q}, \quad (8)$$

where $\rho = 1000\,kg/m^3$, $c_p = 4.18 \cdot 10^3\,J/kgK$ and $k_f = 0.6\,J/mKs$ correspond to the fluid mass density, specific heat (at constant pressure) and thermal conductivity of the fluid, respectively, and $D/D\tilde{t}$ is the material derivative. Here, we have neglected the viscous dissipation term, which is substantially smaller than the Joule heating term [13]. An order of magnitude analysis of (8) gives the diffusion time over which thermal equilibrium is established as $\tilde{t}_{diff} = H^2/\alpha_T \approx 0.14 s$, where $\alpha_T = k_f/\rho c_p = 1.43 \cdot 10^{-7} \, m^2/s$ is the thermal diffusivity and $H = 140 \mu m$ the channel depth over which the largest temperature gradients exist. Thus it is evident that for sufficiently large frequencies (i.e. $\tilde{\omega} \gg \tilde{\omega}_{diff} = 2\pi/\tilde{t}_{diff} = 7.3 Hz$) the differential temperature change, $\Delta\tilde{T}/\tilde{T} \approx \tilde{\omega}_{diff}/\tilde{\omega}$, will be negligible. Hence, for $\tilde{t} > \tilde{t}_{diff}$ and applied frequencies $\tilde{\omega} \gg \tilde{\omega}_{diff}$ steady-state conditions can be assumed while solving for the time-averaged temperature field. Additionally, we note that the heat convection term is negligible relative to the diffusion term in Eq.(8), i.e. $\left|\rho c_p \tilde{\mathbf{u}} \cdot \tilde{\nabla}\tilde{T}\right| \ll \left|k_f \tilde{\nabla}^2 \tilde{T}\right|$. Their ratio can be shown to scale as $\rho c_p u H^2/k_f L$ and for our system, this value is always much smaller than one - even in the case of a relatively large characteristic velocity of $u \approx 1 mm/s$ and channel length $L = 15 mm$, $\rho c_p u H^2/k_f L \approx 0.01$. Thus, the effect of fluid flow on the temperature field can be neglected and eq.(8) can be simplified to

$$k_f \tilde{\nabla}^2 \tilde{T} + \dot{q} = 0. \quad (9)$$

Unlike momentum and species transport analysis, which is confined to the fluidic domain, thermal analysis presents a unique challenge as thermal diffusion necessarily extends the simulation domain from the region of interest (i.e. the fluidic domain) to encompass a significant portion, if not the entire chip. The consideration of the thermal diffusion process through the solid however, poses significant computational problems as it introduces larger length and time scales as well as three-dimensionality. In order to circumvent this and avoid the "whole chip" approach [19] in which the thermal analysis is performed on the entire 3D chip structure (solid walls and fluidic channel), we will formulate an equivalent 2D model in terms of $\tilde{T} = \tilde{T}(x,y)$ (instead of $\tilde{T} = \tilde{T}(x,y,z)$), in which heat transfer in the z-direction and within the channel walls is included via an effective resistance at the top and bottom substrates. Using an infinitesimal cuboid control volume $(dx,dy,H)$ the following 2D version of the heat equation (8) can be derived

$$k_f\left(\frac{\partial^2 \tilde{T}}{\partial \tilde{x}^2} + \frac{\partial^2 \tilde{T}}{\partial \tilde{y}^2}\right) + \dot{q} - \frac{(\tilde{T} - \tilde{T}_\infty)}{H}\left(\frac{1}{R_{US}} + \frac{1}{R_{BS}}\right) = 0, \quad (10)$$

where $\tilde{T}_\infty = 24°C$ is the environment temperature, $R_{US} = t_{US,PDMS}/k_{PDMS} + 1/h = 0.12 m^2 KW^{-1}$ is the thermal resistance (per unit area) of the upper substrate (US) consisting of a conduction $(t_{US}/k_{PDMS})$ and convection $(1/h)$ resistances in series, while $R_{BS} = t_{BS,PDMS}/k_{PDMS} + t_{BS,Glass}/k_{Glass} = 11.7 \cdot 10^{-4} m^2 KW^{-1}$ is the thermal resistance (per unit area) of the bottom substrate (BS) consisting of two conduction resistances in series. The thickness of the various layers (see Fig.1) are $t_{US,PDMS} = 4mm$, $t_{BS,PDMS} = 30\mu m$, $t_{BS,glass} = 1mm$, with the corresponding thermal conductivities $k_{PDMS} = 0.18 Wm^{-1}K^{-1}$ [19] and $k_{glass} = 1 Wm^{-1}K^{-1}$ [13]. Following [19], a natural convection heat transfer coefficient of $h \sim 10 Wm^{-2}K^{-1}$ is used based on the classical configuration of a "heated upper plate". This value is within the range of typical natural heat convection coefficients $2-25 Wm^{-2}K^{-1}$ [21], all of which would result in $R_{US} \gg R_{BS}$. Since these thermal resistances are connected in parallel (Eq.(10)) the conduction resistance through the bottom substrate $R_{BS}$ is dominant.

Note that in contrast to the upper surface where heat transfer by convection was considered, an isothermal condition is implemented at the bottom surface since it is in contact with the microscope stage which can be assumed to be at a constant temperature. Since most of the steady state heat rejection is through the lower substrate [19] instead of the chip side walls, we assume that

$$(\tilde{T} - \tilde{T}_\infty)/R_{SW} = -k_f \partial \tilde{T}/\partial \tilde{n}, \quad (11)$$

where the resistance (per unit area) of the microchannel side walls is approximated $R_{SW} \approx R_{BS}$ and $\tilde{n}$ is the coordinate normal to the wall pointing into the wall domains.

### C. Hydrodynamics

*Governing equations*

The time-averaged velocity of the fluid **u** is governed by the Stokes equation in the low Reynolds number limit ($\rho u H/\mu \sim 0.1$ for typical values for velocity $u \sim 1 mm/s$ and dynamic viscosity $\mu \sim 10^{-3} kgm^{-1}s^{-1}$)

$$-\tilde{\nabla}\tilde{p} + \mu\tilde{\nabla}^2\tilde{\mathbf{u}} + \langle \tilde{\mathbf{f}}_E \rangle = 0, \quad (12)$$

together with the mass conservation equation for an incompressible fluid

$$\tilde{\nabla} \cdot \tilde{\mathbf{u}} = 0, \quad (13)$$

where $\tilde{p}$ is the pressure and $\langle \tilde{\mathbf{f}}_E \rangle$ is the electric force density *outside* the EDL. The effect of $\langle \tilde{\mathbf{f}}_E \rangle$ *within* the thin EDL is captured in the following slip velocity boundary condition.

*Hydrodynamic boundary conditions*

The fluid velocity **u** satisfies the no-penetration condition and the quasi-steady Helmholtz-Smoluchowski slip condition [22]

$$u_{slip} = \mathbf{u} \cdot \mathbf{t} = -\mathrm{Re}\{\zeta\} \mathrm{Re}\{\mathbf{E}_\parallel\}, \quad (14)$$

which is strictly valid for frequencies much smaller than that corresponding to the viscous relaxation within the EDL ($\lambda^2/\nu \approx 0.1 ns$ where for aqueous solutions we take the kinematic viscosity $\nu \approx 10^{-2} cm^2 s^{-1}$ and $\lambda \sim 10 nm$). Here the slip velocity is normalized by $\varepsilon_0 \varepsilon_f E_0 \phi_0 / \mu$, **t** is the unit vector tangential to the wall, $\mathbf{E}_\parallel = \mathbf{E} \cdot \mathbf{t}$ is the component of the electric field tangential to the wall in the fluid domain and $\mathrm{Re}\{...\}$ stands for the real part of $\{...\}$. The induced zeta potential is derived from (3) as

$$\zeta = \underline{\zeta} e^{j\omega t} = \left( \underline{\phi}_w - \underline{\phi}_f \right) e^{j\omega t}. \quad (15)$$

Thus the time-averaged slip velocity is

$$\langle u_{slip} \rangle = -\frac{|\underline{\zeta}||\mathbf{E}_\parallel|}{2} \cos\left( \measuredangle \underline{\zeta} - \measuredangle \mathbf{E}_\parallel \right), \quad (16)$$

with $|...|$ being the absolute value and $\measuredangle$ the phase angle of the complex amplitudes.

*Electrothermal body force*

To first order approximation (obtained by performing a perturbative expansion assuming the relative changes in the permittivity, $\varepsilon$, and conductivity, $\sigma$, are small for typical temperature increments), the time averaged electrical force density for harmonic AC forcing, with the electric field expressed as $\tilde{\mathbf{E}} = \underline{\tilde{\mathbf{E}}} e^{j\tilde{\omega}\tilde{t}}$ is ([13], [20])

$$\langle \tilde{\mathbf{f}}_E \rangle = \frac{1}{2} \mathrm{Re} \left( \frac{\sigma \varepsilon (A-B)}{\sigma + j\tilde{\omega}\varepsilon} \left( \tilde{\nabla}\tilde{T} \cdot \underline{\tilde{\mathbf{E}}} \right) \underline{\tilde{\mathbf{E}}}^* - \frac{1}{2} \varepsilon A |\underline{\tilde{\mathbf{E}}}|^2 \tilde{\nabla}\tilde{T} \right), \quad (17)$$

where $A = \frac{1}{\varepsilon}\left(\frac{\partial \varepsilon}{\partial T}\right)$, $B = \frac{1}{\sigma}\left(\frac{\partial \sigma}{\partial T}\right)$ and * indicates complex conjugate. For an aqueous KCl solution $A \approx -0.4\%K^{-1}$ and $B \approx 2\%K^{-1}$ [13]. However, within this work, we chose to treat $A, B$ as fitting parameters between the numerical and experimental results (see Results and Discussion section). The force density is frequency dependent and has two distinct limits. At low frequencies ($\tilde{\omega} \ll \sigma/\varepsilon$) the Coulomb force (1$^{st}$ term of the right hand side of (17)) dominates since, for an aqueous solution, the relative change in the conductivity is greater than that of the permittivity (i.e. $|B| > |A|$). At high frequencies, $\tilde{\omega} \gg \sigma/\varepsilon$ and the dielectric force dominates. For a typical experimental value of $\sigma \sim 2\,\mu S/cm$ with $\varepsilon = \varepsilon_0 \varepsilon_f$ wherein $\varepsilon_0 = 8.854 \cdot 10^{-12} CV^{-1}m^{-1}$ is the permittivity of vacuum and the relative dielectric constant of water is $\varepsilon_f \sim 80$, one obtains $\sigma/\varepsilon \sim 0.3 MHz$.

**D.     Summary of equations**

Following the above set of approximations, the electrical, thermal and mechanical problems are *decoupled* and we solve the governing equations as follows. First we solve the electrical problem consisting of the Laplace equation (1) for the potentials in both the fluid and wall domains, $\tilde{\phi}_f$ and $\tilde{\phi}_w$ respectively, with the boundary condition (2a,b) to relate them. Second we calculate the temperature field $\tilde{T}$ in the fluid (eqs.(7), (10) and (11)) using the solution for the electric field $\tilde{\mathbf{E}} = -\tilde{\nabla}\tilde{\phi}_f$. Finally, we use these two solutions to compute the electrical force density (17) and solve the Navier–Stokes' equation (12-13) for the fluid velocity, $\tilde{\mathbf{u}}$, and pressure, $p$ with the ICEO slip velocity boundary condition (16).

In order to facilitate the numerical simulations we have chosen to resolve a computational domain in the area of interest that is much smaller than the full chip dimensions (Fig.1), wherein the lengths of the narrow (width $h_1 = 80\,\mu m$) and wide (width $h_2 = 400\,\mu m$) are taken as $l_1 = 1.6mm$ and $l_2 = 4mm$, respectively, instead of $L = 15mm$. The electric potential drop between the opposite microchannel entrances can then be approximated by $\Delta V' \sim \Delta V\left((l_1 h_2 + l_2 h_1)/(Lh_2 + Lh_1)\right)$, which in the case of an applied potential drop of $\Delta V = 2000V$ yields $\Delta V' \approx 266V$. A more accurate value of $\Delta V' \approx 274V$ is obtained through solving the Laplace equation $\nabla^2 \underline{\phi}_f = 0$ in the full chip geometry, which accounts for the curvature of the corner, with an electric potential drop of $\Delta V = 2000 V$ between the microchannel ends with an insulation condition at all other solid-electrolyte interfaces. The radius of curvature of the L-junction corner used in the simulations was 6μm.

## IV. RESULTS AND DISCUSSION

*Temperature distributions for different solution conductivities*

Based on the measured fluorescence intensity (see supporting materials [23]), Figure 2 illustrates that the average temperature within the fluid increases with increasing solution conductivity [24] since the Joule heating effect is linearly dependent on the medium conductivity ($\dot{q} \propto \sigma E_0^2$). More specifically, the time evolution of the averaged normalized temperature within the narrow (Fig.2b) and wide (Fig.2c) channels shows a sharp increase (decrease) upon application (shut off) of the electric field at t=10s (t=30s). It is also clear that the Joule heating effect is more pronounced within the narrow channel compared to the wide one, in accordance with the existence of larger electric fields within the former. The 2D intensity maps at time t=30s (just before shut off) clearly show the existence of sharp axial temperature gradients at the vicinity of the channel corner, whereas an almost uniform temperature distribution exists along the channel's axial direction away from the corner. This is a clear indication that most of the generated heat dissipates orthogonally to (rather than along) the channel's axial direction, thus, supporting the main assumption of the above thermal theoretical analysis, i.e., the consideration of heat dissipation through the top and bottom surfaces of the microfluidic channels (3$^{rd}$ term on the r.h.s. of eq.(10)) as well as through the side walls (boundary condition eq.(11)).

An order-of-magnitude estimate of the incremental temperature rise (within the narrow channel) can be made based on eqs. (7) and (9) as $k_f \frac{\Delta \tilde{T}}{H^2} \approx \sigma \left( \frac{\Delta V}{L(1+h_1/h_2)} \right)^2$ or

$$\Delta \tilde{T} \approx \frac{\sigma}{k_f} \left( \frac{\Delta V \cdot H}{L(1+h_1/h_2)} \right)^2,$$ where $\Delta V$ is the potential difference across the electrodes, $h_1 = 80 \mu m$ and $h_2 = 400 \mu m$ are the narrow and wide channels widths, respectively. For an applied potential amplitude of $\Delta V = 2000 V$ at 300Hz and with $\sigma = 2, 50, 350 \, \mu S/cm$ the temperature rise can be approximated as $\Delta \tilde{T} \approx 0.1, 2, 14 °C$, respectively. This stands in reasonable agreement with the experimental results in Figure 2.

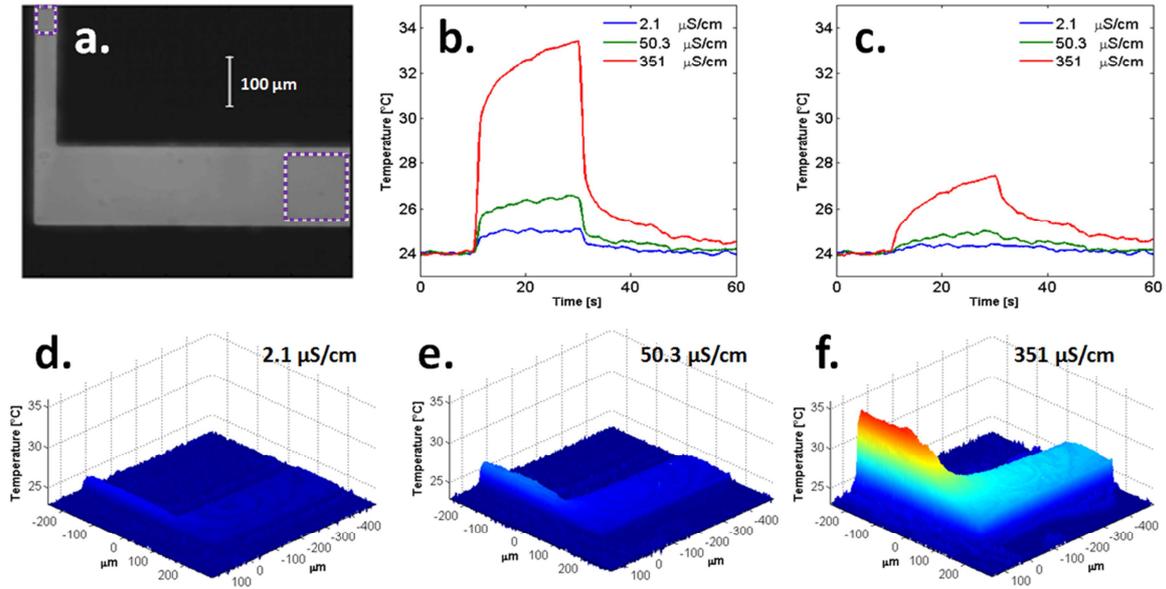

**Figure 2: (color online)** Experimental visualization of the temperature field using Rhodamine B. (a) The interrogation windows used for generating the time evolution of the average temperature at the narrow (b) and wide (c) channels at various solution conductivities. The applied voltage amplitude of 2000V and 300Hz was introduced at t=10s and shut off at t=30s. Two-dimensional temperature maps at t=30s (just before shutting off the voltage) for various solution conductivities: (d) 2.1μS/cm; (e) 50.3μS/cm; and (f) 351μS/cm.

The numerically calculated temperature distribution that is shown in Fig.3 qualitatively supports the experimental findings (Fig. 2), such that indeed most of the axial temperature gradients exist at the vicinity of the corner along with elevated temperatures within the narrow channel. The latter are in good quantitative agreement with the experimental observations (Fig.2). The magnitude of these temperature gradients implies that they cannot be neglected and will give rise to a significant electrothermal body force (Fig.3(b)), which in turn, affects the local flow pattern as discussed in more detail in the following sections.

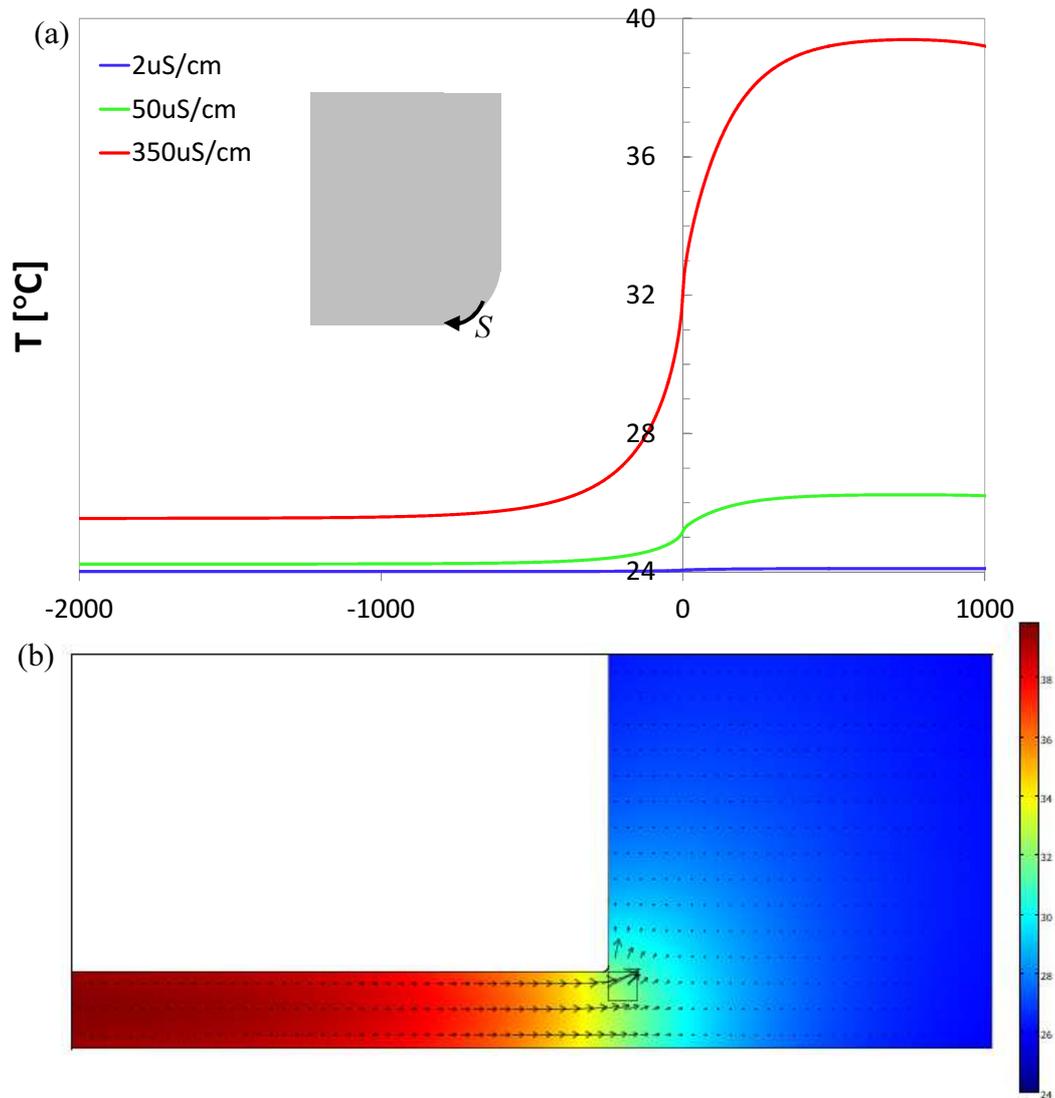

**Figure 3: (color online)** Numerical simulation results at a fixed frequency of 300Hz and 2000V applied voltage showing the: (a) temperature profiles for various conductivities along the corner wall versus the coordinate $S$ (see inset); (b) electro-thermal body force (vector plot) and temperature (surface plot) distributions for 350μS/cm.

*Flow patterns for different solution conductivities*

The μPIV generated velocity fields for solutions of varying conductivity, at a fixed voltage (2000V) and frequency (300Hz) are shown in Fig.4. In contrast to previous studies ([6], [7]) on a similar device but performed under DC field conditions, where a superposition of linear electroosmotic and ICEO flows must be accounted for, here only the ICEO contribution is seen as the linear contribution is time-averaged to zero. For a solution of low conductivity ($2\,\mu S/cm$), the flow field clearly exhibits the expected non-linear ICEO jetting flow, stemming from the polarized induced zeta-potential around the corner (Fig.5), along with a vortex pair that is formed due to conservation of mass[5]. The maximum value of the induced zeta potential at the largest applied voltage, 2000V in amplitude is ~10mV, hence, smaller than the thermal potential.

As the conductivity of the solution is increased, a decrease in the calculated effective zeta potential is observed, which corresponds with previous experimental results [25]. At the same time, a clear divergence of the flow pattern from the expected ICEO jetting is obtained. The onset of this divergence corresponds with the increase in temperature gradients within the electrolyte as illustrated in Figure 2. By incorporating electrothermal effects in the numerical solutions (Fig.4 (d)-(f)) the obtained flow patterns are in good qualitative agreement with the experimental observations. In the sequel, we further assess the influence of Joule heating, by examining the voltage and frequency dependent behavior of electrolytes of varying conductivity.

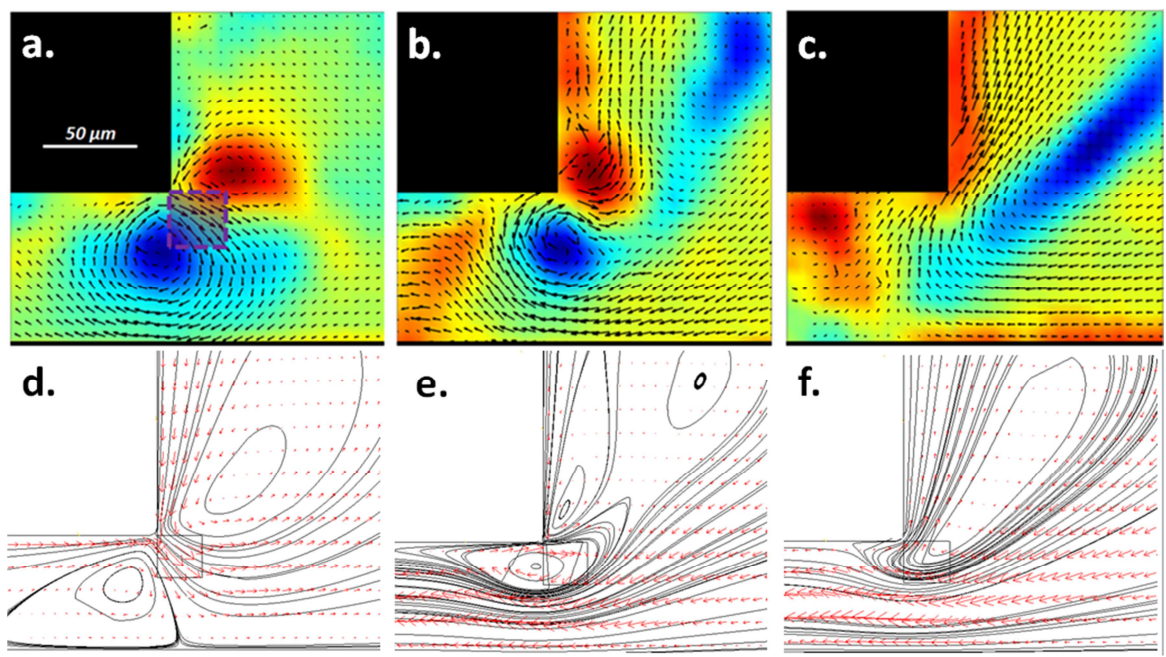

**Figure 4**: **(color online)** (a)-(c) μPIV generated flow fields for a 2000V voltage amplitude, 300Hz signal, and varying solution conductivities exhibit divergence from the expected ICEO corner ejection flow with increasing conductivity. Arrows: velocity vectors. Color: angular momentum parameter - swirl intensity [26]; (d)-(f) The resulting hydrodynamic flow patterns

(velocity streamlines) for various electrolyte conductivities indicating a clear divergence from the classical ICEO ejection flow with increasing conductivity. Arrows: velocity vectors.

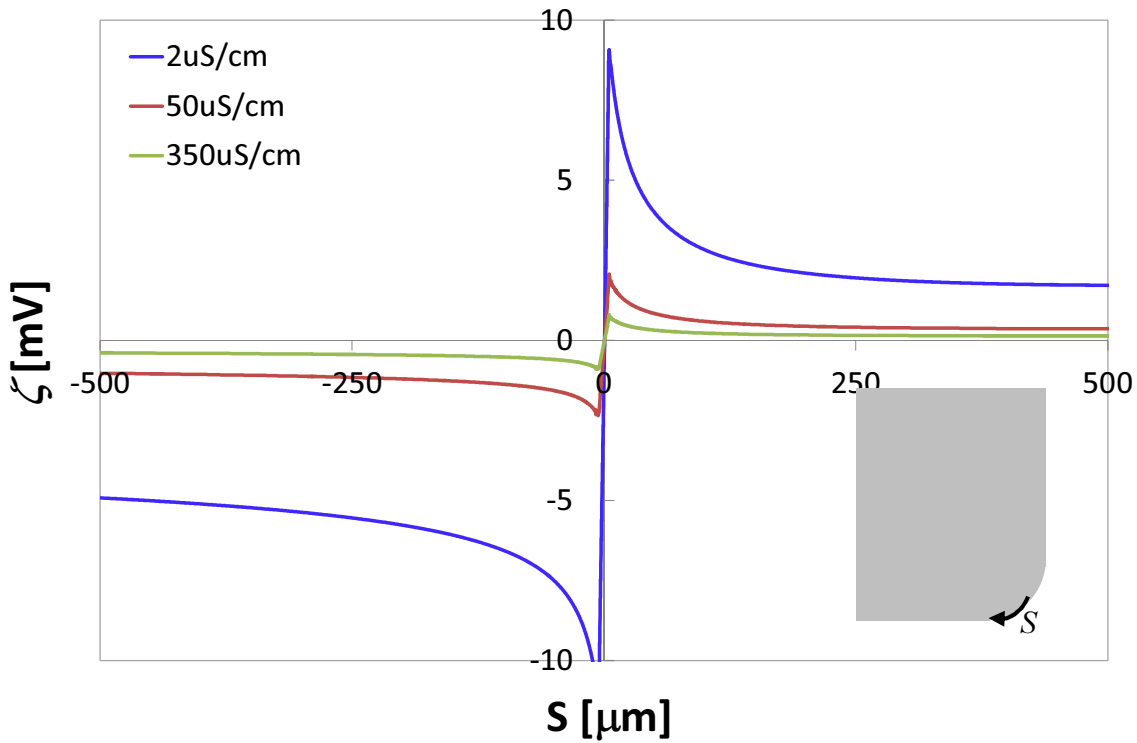

**Figure 5**: **(color online)** Numercially calculated induced zeta-potential around the corner (inset) for a 2000V voltage amplitude, 300Hz signal, and varying solution conductivities.

*Voltage scan*

Fig.6a shows that the measured ejection velocity at the lowest conductivity is quadratic in the electric field at a fixed frequency. This stands in agreement with the theoretical predictions of ICEO phenomenon [8] and previous experimental studies [6]. The measured velocity is obtained by averaging the μPIV velocities in an interrogation box of 30μm X 30μm in size (see Fig.4a), where we differentiate between its absolute magnitude and its component projected on the corner bisector (with its positive sense pointing into the electrolyte). As seen for $\sigma = 2.21\,\mu S/cm$ both experimental and calculated values collapse onto the line of purely ICEO response (i.e. without electro-thermal effects). This result verifies that indeed the ICEO flow consists of strong ejection flow along the corner bisector and that at such low conductivity electro-thermal effect is negligible. It is noted that in order to obtain an agreement between the numerical solution and the experiments, a fitted value for the effective wall dielectric constant of $\varepsilon_w = 43$ (see Fig.S2 in supporting materials [23]) is used rather than the material property for PDMS of $\varepsilon_w = 3$. The need for an artificially increased wall permittivity, implies that the PDMS wall behaves as if it is more polarizable than its material property suggests. The underlying mechanism behind why the numerical simulations underestimate the experimental results is not clear and it is in opposite trend from what is commonly reported in the literature regarding ICEO over conductors (or ideally polarizable), wherein the theoretical and numerical results tend to overestimate the experiments by at least an order of magnitude [25], [27], [28]. Several physical mechanisms have been suggested to address the latter, such as surface conduction [27], non-linear capacitance and steric effects associated with the induced-charge [29], [30]. However, all of these mechanisms add corrections to the numerical solution so as to decrease the resulting ICEO velocities and hence are likely not relevant in the current study wherein we are facing an opposite problem of how to increase the numerically predicted velocities. Extension of the weak field approach ($\phi_0 \ll RT/zF$), underlying eqs.(2a,b), to strong fields ($\phi_0 \sim RT/zF$) about a dielectric surface of zero surface charge [31] also cannot explain this discrepancy, and on the contrary, shows a transition from an $E_0^2$ dependence at moderate fields to an essential linear variation with $E_0$ at strong fields.

In line with the decreased induced zeta potential (Figure 5) and the departure of the flow pattern from classical ICEO ejection (Figure 4) observed with increased conductivity, we also see a marked divergence between the absolute value of the measured velocity and its projected component. In fact, for $\sigma = 373\,\mu S/cm$, the projected component even reverses sign and instead of pointing outwards from the corner, it points inward. At the same time, although at low voltages the absolute velocities are smaller than the case of $\sigma = 50.6\,\mu S/cm$, at large enough voltage, they exceed even the velocities of the $\sigma = 2.21\,\mu S/cm$ solution. Overall, for $\sigma = 373\,\mu S/cm$, the scaling of the velocity with voltage changes from quadratic to biquadratic (i.e. to the fourth power) as shown in the inset of Fig.6c, in agreement with the theoretical scaling of the electro-thermal effects [32]. The dominance of the electro-thermal effects at high conductivity solutions, which are amplified at high applied voltages are further verified by the numerical simulations.

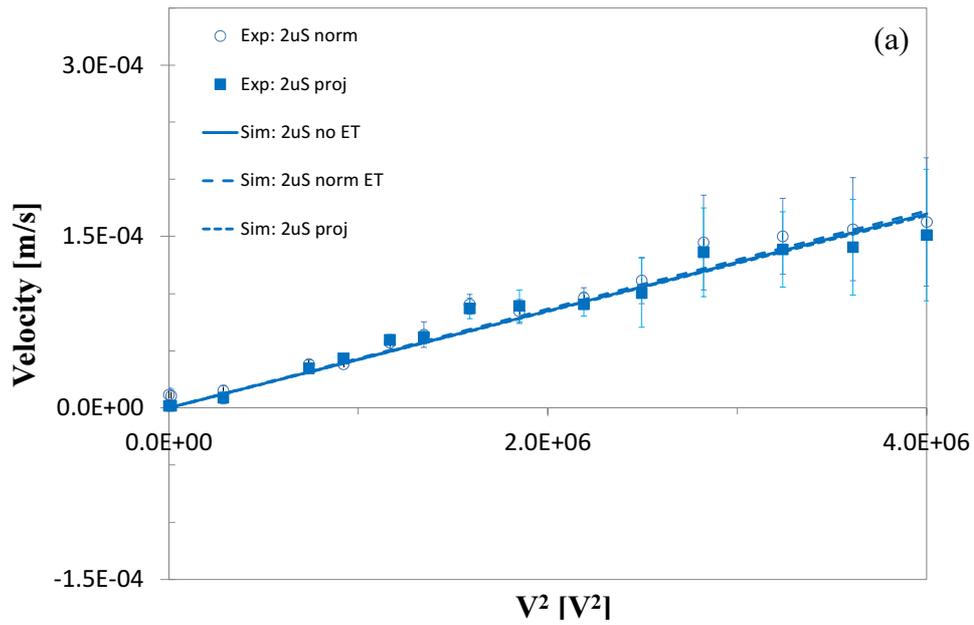
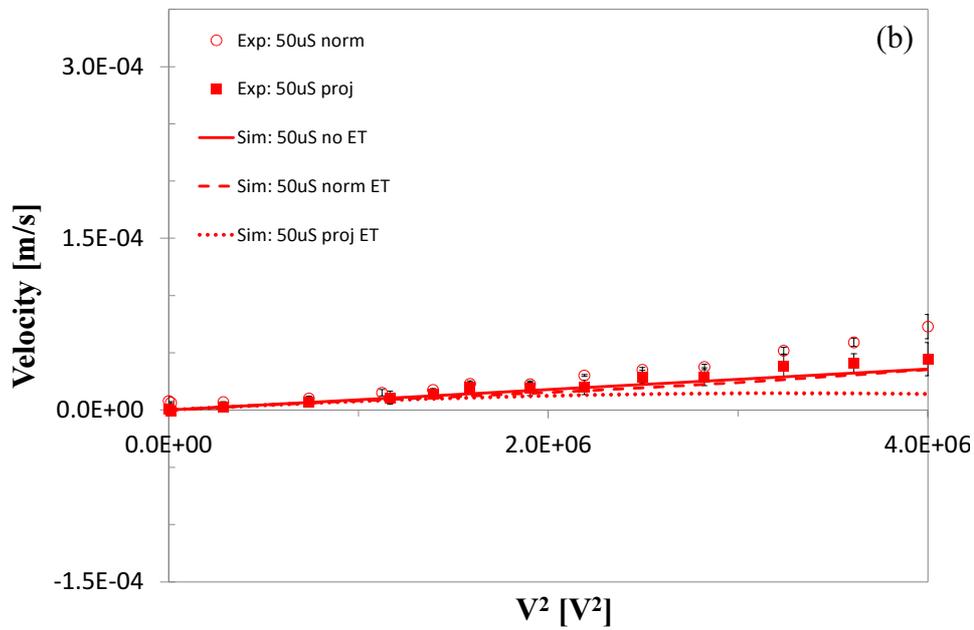
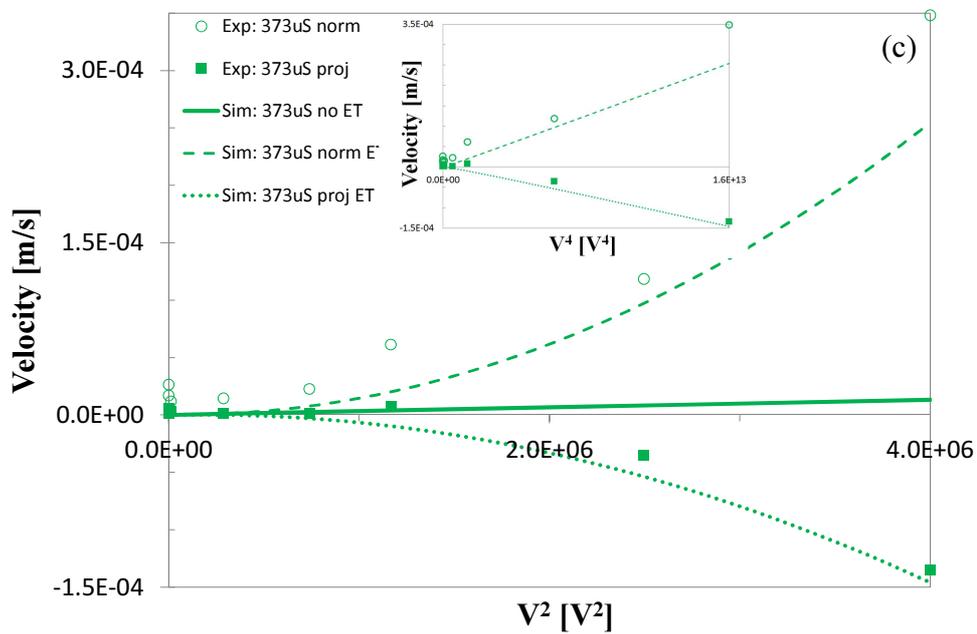

**Figure 6**: **(color online)** Measured (symbols) and calculated ejection velocity versus increasing electric potential amplitude (squared) for different solution conductivities at 300 Hz. Continuous lines stand for the numerical model without electro-thermal (ET) effects; dashed line stand for the norm of the velocity; and dotted line stand for the projected (along the corner bisector) component of the velocity. As seen there is an increasing divergence between the norm and projected values with increasing conductivity. A fourth power dependence of the velocity on voltage is shown in the inset of part (c) indicating the existence of strong electro-thermal effects. Error bars indicate the measured standard deviation.

*Frequency scan*

Here we concentrate on studying the frequency dependence of the ICEO ejection velocity. Based on Fig.6 we use the component of the velocity projected onto the corner bisector to differentiate between ICEO and electro-thermal effects (see also supporting materials [23]). Figure 7 depicts the dependency of the ejection velocity on the applied frequency and exhibits for the case of low conductivity solution (2μS/cm) both low and high frequency dispersion with a maxima around 200-400Hz. Applying a numerical model, which considers the wall to be a lossless dielectric with the fitted effective wall dielectric constant of $\varepsilon_w = 43$, we observe similar high frequency dispersion behavior as in the experiments but also record an order of magnitude difference between the experimental and numerical RC times for the low conductivity solution. The greater discrepancy of two orders of magnitude in RC time for the intermediate conductivity case is likely also influenced by the non-negligible electro-thermal effects.

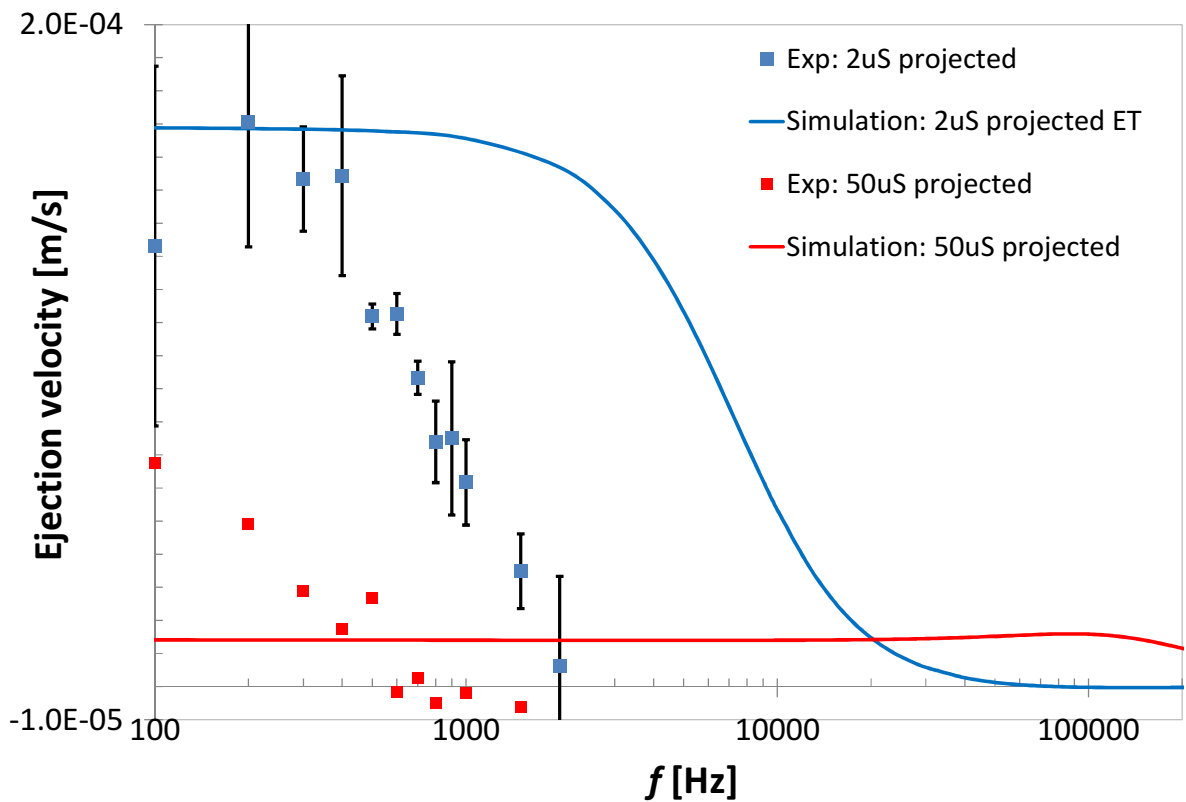

**Figure 7**: **(color online)** Ejection velocity versus frequency as measured (symbols with error bars of a single standard deviation) and numerically calculated (continuous lines) for the case of low conductivity (2μS/cm) solution and applied voltage amplitude of 2000V. Both measured and calculated ejection velocities are the projected velocity along the corner angle bisector averaged within an interrogation window of 30X30μm (see Fig.4a).

V. CONCLUSIONS

As opposed to their DC counterpart [7], the use of AC fields facilitates a direct experimental observation of the non-linear ICEO corner jetting flow by eliminating the linear EOF effects (which have a zero time-average contribution). The ICEO flow pattern was observed using μPIV, clearly showing the corner ejection flow along with the associated vortex pair in excellent qualitative agreement with the theoretical predictions. Quantitative analysis of the ejection velocity demonstrated the expected linear scaling with the square of the applied electric field. In addition both low and high frequency dispersions, associated with the relaxation times of the induced EDLs at the active electrodes and polarized corner, respectively, were observed. Numerical calculations yielded the same high frequency dispersion trend but with an order of magnitude larger RC time. Curiously, we obtained an opposite trend from what is commonly reported for ICEO over conducting (ideally polarizable) medium. Specifically, the numerical solution of the induced velocities underestimated the experimental data by an order of magnitude in the case of a dielectric wall, whereas in the case of conducting surfaces, the numerics tend to overestimate the experiments by at least an order of magnitude. Non-linear physical phenomenon corrections, such as surface conduction, non-linear capacitance and steric effects which tend to decrease the predicted velocity cannot explain this trend and thus the resolution of this discrepancy is left for future study.

The strong divergence of the hydrodynamic flow pattern (from the classical ICEO ejection) with increasing electrolyte conductivity is proven to be a result of electro-thermal effects due to Joule heating by several means. Experimentally, two-dimensional mapping of the temperature distribution, using Rhodamine B, is used to illustrate an overall increase in solution temperature with increasing conductivity. Additionally, it is demonstrated that since the electric field within the narrow channel is larger than that in the wide channel, the former exhibits higher temperatures. These results were in excellent agreement with the numerical solution of the 2D heat equation we have formulated (Eq.(10)) which accounts for heat losses through all electrolyte-solid wall interfaces. Furthermore, accounting for the electro-thermal body force in the numerical simulations resulted in quantitative agreement with the experimental data in terms of both the flow patterns and the velocity dependence on the voltage including the shift from a quadratic to biquadratic scaling with increasing dominance of the electrothermal effects. Since the electro-thermal effects are arise from temperature gradients, which in turn stem from the electric field gradients, suppression of the latter (e.g. uniform channels and/or rounding of the corner) can minimize these electro-thermal effects. Thus, the current study besides providing, an in-depth description of the competition between ICEO and electrothermal effects, can also be used to guide the design of planar microfluidic systems.

**ACKNOWLEDGMENTS**

This work was supported by ISF Grant No. 1078/10. We thank the Technion RBNI (Russell Berrie Nanotechnology Institute) for their financial support.